\pdfoutput=1

\documentclass[multi]{EngC}

\usepackage[unicode=true,%
 pdftitle=Analysing Shortcomings of Statistical Parametric Speech Synthesis (Draft Book Chapter),%
 pdfauthor={Gustav Eje Henter, Simon King, Thomas Merritt, Gilles Degottex},%
 bookmarks=true,bookmarksnumbered=true,bookmarksopen=false,%
 breaklinks=true,pdfborder={0 0 0},backref=false,colorlinks=true,citecolor=blue]{hyperref}

\usepackage[rightcaption,raggedright]{sidecap}
\usepackage{framed}         
\usepackage{soul}           

\usepackage[agsm]{harvard}

\usepackage{rotating}
\usepackage{floatpag}
\rotfloatpagestyle{empty}

\usepackage{amsmath}
\usepackage{amssymb}
\usepackage{amsthm}
\usepackage{bm}
\usepackage{txfonts}
\usepackage{graphicx}
\usepackage{subfigure}
\usepackage{multirow}

\usepackage{index}
\newindex{aut}{adx}{and}{Author index}
\makeindex

\usepackage{xspace} 

\usepackage{comment}

\theoremstyle{plain}

\theoremstyle{definition}

\theoremstyle{remark}

\newcommand{\ignore}[1]{}
\newcommand{\todo}[1]{}
\newcommand{\notes}[1]{}

\begin{document}
\title[Draft Book Chapter]{Analysing Shortcomings of Statistical Parametric Speech Synthesis}

%


\thispagestyle{empty}
~\vfill
\noindent{}\textit{This is a draft chapter for the book ``Natural Speech Technology: Statistical and Machine Learning Approaches to Speech Recognition and Synthesis''.}

\textit{The chapter text was mainly written prior to and during November 2016, and reflects the shape and general understanding of statistical speech synthesis at that time. The draft was edited in July 2017 for length and edited again in July 2018 to function as a stand-alone document, as presented here. This version was compiled on \today.}
\vfill
\newpage






\author{Gustav Eje Henter, Simon King, Thomas Merritt\hyperref[amazon]{\textsuperscript{\textdaggerdbl}} and Gilles Degottex}

\chapter{Analysing Shortcomings of Statistical Parametric Speech Synthesis}
\label{ch:diagnose}


\label{ch18start}
\section{Introduction}
\label{ch18-diagnose_introduction}


\newcommand\blfootnote[1]{%
  \begingroup
  \renewcommand\thefootnote{}\footnote{#1}%
  \addtocounter{footnote}{-1}%
  \endgroup
}

Even\blfootnote{\phantomsection\label{amazon}\textsuperscript{\textdaggerdbl} Work done prior to joining Amazon.} the best statistical parametric speech synthesis (SPSS) produces output that is noticeably worse than natural speech, whether measured in terms of quality, naturalness, speaker similarity, or intelligibility in noise. These \emph{shortcomings}  have been attributed to a multitude of causes, and the literature is awash with solutions to various supposed problems with the statistical parametric approach. Yet, somewhat surprisingly, the hypothesised problem is often not clearly defined, or no empirical evidence us provided to confirm -- or quantify -- its contribution to imperfections in the synthetic speech. Across the literature as a whole, there is surprisingly little work exploring which of the many potential problems are perceptually the most important; this would, of course, be useful knowledge.

The conventional arguments in favour of the statistical parametric approach are, in contrast, well rehearsed and \emph{are} supported by plenty of experimental evidence: good intelligibility, robustness to imperfect data, ability to adapt the model, and so on.

If we take, as is widely accepted in the field, \emph{naturalness} as our principal measure, then evidence that synthetic speech is inferior to natural speech recordings is overwhelming. The simplest and clearest evidence comes from the the long-running Blizzard Challenge, summarised in \cite{king2014measuring}. Year after year, we see that no synthesiser is ever judged to be as natural as recorded speech.

In contrast to naturalness, the Blizzard Challenge shows that impressive progress has been made in \emph{intelligibility}, where -- in quiet conditions at least -- some synthesisers are as good as recorded speech. This is not (yet) generally the case in non-quiet listening situations. The Hurricane Challenge \cite{cooke2013intelligibility} provides convincing evidence that synthetic speech intelligibility in the presence of additive noise remains substantially inferior to that of recorded speech.

The Blizzard and Hurricane Challenges both compare different synthesis approaches on the same data, demonstrating that it is the specific implementation and assumptions of each synthesiser that are responsible for differences in naturalness or intelligibility. There are apparently only two or three different waveform generation technologies represented in the entries the these challenges: statistical parametric systems employing a vocoder, unit selection employing waveform concatenation, and a \emph{hybrid} variant of unit selection that uses an internal statistical parametric model. There are clear trends, consistently across many years of the challenge: 1.\ statistical parametric systems are generally the most intelligible; 2.\ unit selection systems are generally the most natural-sounding; 3.\ hybrid systems can achieve the naturalness of unit selection whilst approaching the best intelligibility. Strictly speaking, we can only say say that the choice of waveform generation technology \emph{correlates} with intelligibility and naturalness; we can't make the stronger claim of \emph{causality}.

Because individual system details are seldom open knowledge, and because entire systems are evaluated ``end to end'', it is not straightforward to attribute successes and shortcomings to specific elements in each approach. This is particularly true about the front-end text processor, partly because much recent research has neglected the effect of this component and focussed much more on acoustic modelling and waveform-generation technology.


In this chapter, we analyse some of the shortcomings of SPSS. Since there is at least a correlation between waveform generation technology and synthetic speech naturalness, we start with vocoding. This is followed by a description -- with an example application -- of a general methodology for quantifying the effect of any of the many assumptions, design choices, and (possibly inherent) limitations of SPSS. The example application includes measuring the shortcomings of vocoding, relative to other limiting factors such as the statistical model.


\section{Vocoding}
\label{ch18-diagnose_vocoding}

The role of the vocoder in SPSS is to provide a representation of the speech signal that is suitable for statistical modelling and at the same time from which a waveform can be generated. Therefore, vocoder design inevitably involves a trade-off between the two. For example, dimensionality reduction and an approximately decorrelating transform, e.g., retaining only low-order mel-cepstrum coefficients, are widely applied, either within the vocoder, or to its parametrisation.

Vocoded (analysis-synthesis) speech is the assumed upper-bound of the quality achievable from a SPSS system in the limit of a highly accurate statistical model, and it is therefore of interest to measure how much the vocoder alone limits the achievable quality of all systems that employ one.

The very act of parametrising a speech signal, which is itself a non-linear combination of interacting sound sources and sound-shaping processes, creates many challenges. For example, it is known that the glottal source has a particular amplitude spectrum which nevertheless most vocoders assume is flat, combining all spectral envelope modelling into a single component that also handles the amplitude spectrum of the Vocal Tract Filter (VTF). However, the true glottal source amplitude spectrum varies with F0. Most current vocoders take a simple approach, assuming the VTF is independent of F0, and therefore with no separate modelling of glottal source amplitude spectrum, non-periodic sound generated in the glottis, or the glottal source phase spectrum.

Another example is the binary voicing decision (speech is either voiced or unvoiced) that many vocoders incorporate, which is arguably an oversimplification. In natural speech, at transitions from unvoiced to voiced, or vice versa, it is frequently the case that the so-called deterministic (i.e., voiced, periodic) speech component somewhat gradually commences or fades away. As a consequence, the way in which voicing is handled by a vocoder may lead to differences in synthetic speech quality \cite{latorrej2011contf0,yukai2011contf0,degottexg2014jhmpd}.

As a final example of the many issues in vocoding, the seemingly random timing of glottal pulses in creaky voice, also called vocal fry \cite{laverj2009vocqual}, is generally poorly captured in current vocoders because they use a perfectly periodic pulse train excitation signal for all voiced sounds. Irregularities in the mechanical vibration of the vocal folds are commonplace \cite{drugmant2014creaky}. Current speech analysis techniques are prone to confusing these irregularities with simple additive noise, which leads to vocoders producing speech with a perceived hoarse voice quality instead of a creaky quality. Similarly, breathy voice \cite{laverj2009vocqual,ishict2010breathy} can only by synthesized by the simple addition of noise, whereas in actual speech production the shape of the glottal pulse might be also varying rapidly, producing a signal that cannot be approximated by a pulse train plus noise.

Even though there are some advanced techniques for creak detection \cite{kanej2013detectcreak,drugmant2014creaky} and measures of pulse variation \cite{degottexg2014jhmpd} it is not obvious how to build a vocoder that takes advantage of such features in a statistical modelling framework, without substantially increasing the dimensionality of the representation.

The standard speech parametrisation setup used in statistical parametric speech synthesis is outlined in \cite{zen2009statistical}. However, for the reasons above, and others, this configuration has a degrading effect on perceived quality of speech \cite{merritt2014investigating,henter2014measuring,merritt2015attributing}. Furthermore, the amount of degradation may differ markedly between speakers (voices), and one vocoder or another may perform better or worse for any particular speakers; cf.\ \cite{babacan2014parametric} for singing. These quality variations suggest a notion of ``vocodability'', the consequence of which is a (perhaps undesirable) bias when selecting a speaker for a TTS corpus, towards a voice that suffers minimal degradation at the hands of the vocoder.

Approaches to improving the modelling of speech signals for SPSS can be described in three distinct categories: source-filter parametrisations, sinusoidal parametrisations and non-parametric approaches. A further comparison of selected source-filter and sinusoidal parametrisation methods can be found in \cite{hu2013experimental}.



\subsection{Source-filter parametrisation}
\label{ch:diagnose:source-filter}

The idea that source and filter can be separated is a simplification of the voice production mechanism. Coupling exists: when articulator positions change, this not only changes the VTF, but also has an effect on the glottal pulse spectrum \cite{fantg1987interact}; the more open the glottis, the wider the formant bandwidths \cite{hanson1997glottchar}; etc. So, source and filter are obviously dependent and correlated, since they are the intertwined consequences of articulator movement. The acoustic consequences of source and of filter are therefore also not entirely separable. We examine the impact of independent modelling of source and filter in section \ref{ch18-diagnose_streams}.

The other obvious limitation of the source-filter model is that the location of the sound source is not always the glottis, but can be elsewhere in the vocal tract, such as the constriction for a fricative, or the closure and release of a plosive. For these reasons, we should state clearly that source-filter models are only models of speech signals, \emph{inspired} by speech production mechanisms, but not faithful to them.

One line of research that may eventually lead to an alternative solution, one that \emph{is} faithful to speech production mechanisms, is so-called physical modelling; that approach is still a long way from offering this solution: the models are simplified and incomplete, and fitting their parameters to natural speech signals cannot be done reliably. Therefore, the vast majority of SPSS systems use models of the speech signal, not of the production process.

Staying within the source-filter signal modelling paradigm, it may be possible to improve modelling accuracy with a more sophisticated source. As mentioned earlier, the glottal source magnitude and phase should ideally not be assumed constant and flat, although this is what STRAIGHT \cite{kawahara2006straight} and most other vocoders assume. Natural glottal pulses actually have a non-minimum-phase spectrum which therefore cannot be predicted from the amplitude spectrum. This is also true of the VTF.

Many approaches have been suggested to improve the model of the glottal source for SPSS \cite{klatt1990,raitio2011hmm,cabral07,degottexg2013svln}
either by estimating the glottal pulse waveform \cite{raitio2011hmm} or the parameters of an analytical model of it \cite{cabral07,degottexg2013svln}. In addition to attempting to solve the issues mentioned above, the analytical model approach also provides parameters that are not merely generic signal-based features, but more closely related to the underlying physical system. For example, the $Rd$ coefficient \cite{fantg1995} links the amplitude and phase spectra of the glottal source in a way that is governed by physical constraints. 




\subsection{Sinusoidal parametrisation}

The other main type of signal analysis used in SPSS is the sinusoidal model \cite{mcaulay1986,stylianouy1996}, although this is most often used only as an intermediate representation, with a subsequent dimensionality-reducing parametrisation being required for statistical modelling. Sinusoidal models can be seen as sparse representations of the speech signal:
\begin{equation}
s(t) = \sum_{k=-K}^K a_k(t)\cdot e^{j\phi_k(t)}
\end{equation}
where $K$ is the number of sinusoidal components, with amplitude $a_k(t)$ and phase $\phi_k(t)$ \cite{degottexg2013jahmair}. The spectral amplitude envelope can be constructed from the amplitude parameters \cite{eljaroudia1991dap} while a phase envelope can be found from the phase parameters \cite{agiosy2009wrapgauss,degottexg2014jhmpd}.

\begin{figure}[!t]
\centering
\includegraphics[width=0.92\columnwidth]{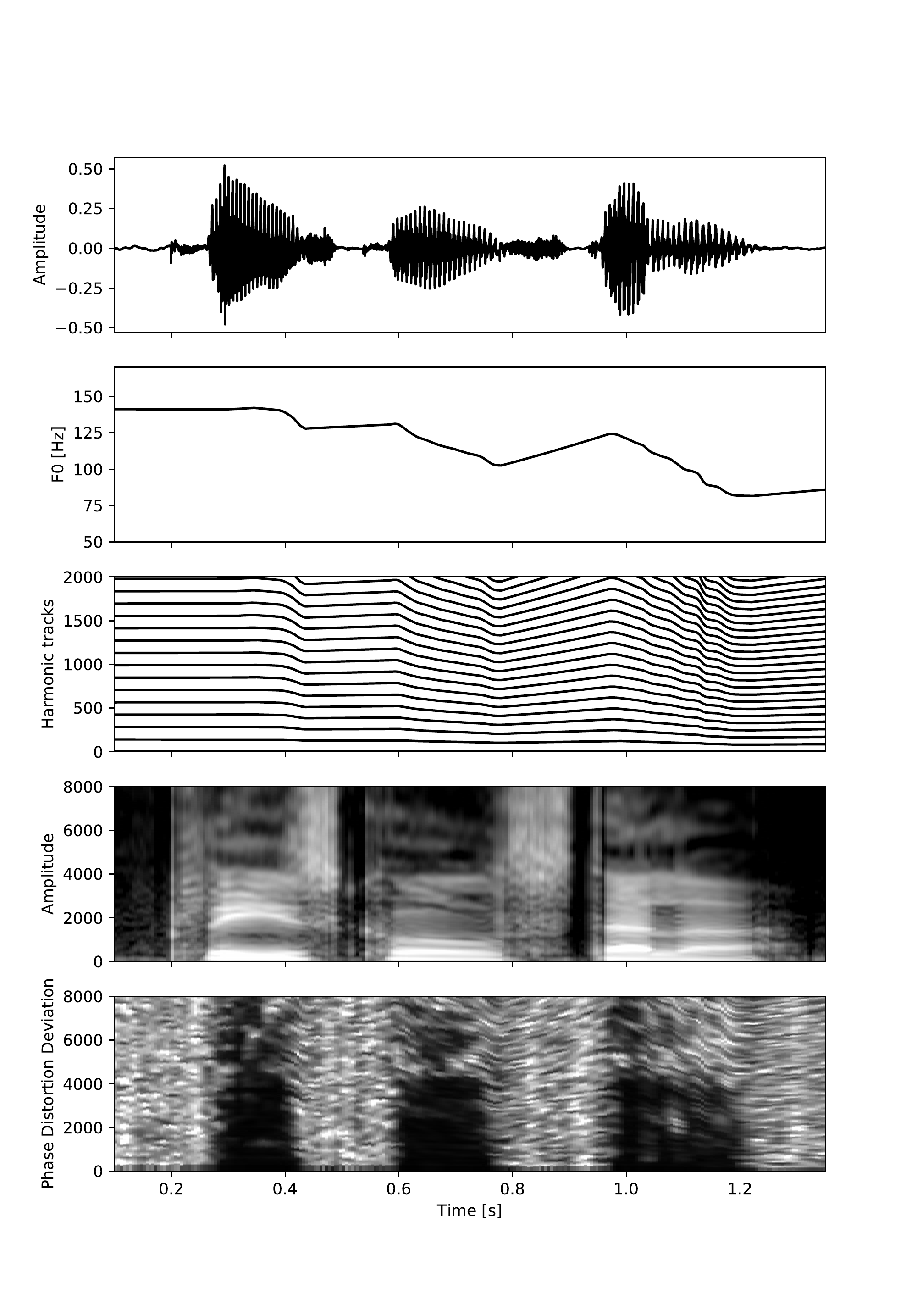}
\vspace{-1cm}
\caption[An example of parameters estimated by the HMPD vocoder.]{An example of parameters estimated by the HMPD vocoder. From top to bottom: waveform; continuous fundamental frequency F0 curve; harmonic tracks used for the estimation of $a_k(t)$ and $\phi_k(t)$; the amplitude spectral envelope modelling $a_k(t)$ (the lighter the colour the louder); the Phase Distortion Deviation (PDD) modelling the random component of $\phi_k(t)$ (the lighter the colour the noisier)}
\label{figure:example_speech_sinparameters}
\end{figure}


Because the reconstruction quality of sinusoidal models is very high \cite{degottexg2013jahmair}, attempts have been made to directly model the parameters \cite{huq2016comvalspss} without requiring further parametrisation (e.g., as a spectral envelope represented by a truncated mel cepstrum). Direct modelling of sinusoidal model parameters falls somewhere between the fully parametric source-filter approach above (e.g. STRAIGHT) and the non-parametric approach described next.

An open issue is how to manage the dimensionality of sinusoidal models \cite{huq2016comvalspss}, since there are a large number of highly-correlated parameters. A further complication is that the number of sinusoids below the Nyquist frequency varies over time.

\subsection{Non-parametric representation}
\label{ch18-diagnose_nonparametric}

Given that the vocoder limits naturalness, there is interest in integrating some or all of the vocoder signal processing into the statistical model, essentially enabling the parametric speech representation to be learned and improved, rather than being static (even if carefully engineered). Automatic Speech Recognition and other signal processing applications have undergone significant changes quite recently to using lower-level features. For speech synthesis, one example is predicting the full-detail, high-dimensional STRAIGHT spectrum rather than working with mel-generalised cepstral coefficients (MGCs), e.g., \cite{ling2013modeling}. This removes a potentially restrictive dimensionality reduction, at the expense of having to learn to model highly correlated features.

Going further, there are ways to directly generate a waveform without reconstructing it from an engineered intermediate representation such as the spectral envelope.
One step towards direct waveform generation was the generation of glottal pulse waveforms glottal source signal in speech \cite{raitio2014voice}, even down to the level of individual samples \cite{juvela2016high}, although these were then used in a source-filter model. Another way to generate a waveform is to use the output from an SPSS system to control the selection of units in a unit selection system: hybrid synthesis, described briefly below.  Attempts have also been made to statistically model the spectral properties directly from waveforms instead of passing through an explicit estimate of the spectral envelope \cite{tokuda2015directly,tokuda2016directly}.

Most recently, WaveNet \cite{oord2016wavenet} suggests that direct waveform synthesis is not merely a theoretically interesting concept, but is capable of very high output quality. This performance currently comes at prohibitive computational cost, as well as concerns over data quantity requirements and patent status, which may limits the usefulness of this particular approach.


An alternative technique to obtain high segmental quality is to use waveform segments from the training database for signal generation. Hybrid synthesis (unit selection driven by SPSS) is well-established and in widespread commercial use. Hybrid systems have made a strong showing in the Blizzard Challenge ever since their arrival \cite{ling2007ustc,ling2008ustc,lu2009ustc,jiang2010ustc}, the approach has subsequently grown in popularity there \cite{king2014measuring}, and Open Source implementations are becoming available \cite{merritt2016deep}. Other permutations of SPSS and unit selection include multiform synthesis \cite{pollet2008synthesis}, where a sequence of SPSS-generated speech and recorded waveform units are concatenated, or manipulating recorded speech prototypes to match predictions from SPSS \cite{espic2016waveform}.

\subsection{Summary}
The near future of speech synthesis will certainly involve continued attempts to move beyond the use of vocoders. As deep learning improves the state of the art in many areas of spoken language processing, including TTS, it becomes more obvious how to replace traditional signal processing with a learned pipeline. By recasting some or all of the acoustic feature processing as layers of neural network (for example), end-to-end optimisation of acoustic model and signal representation becomes apparently straightforward. For example, \cite{takaki2015multiple} investigated the many roles that a DNN might fulfil in a TTS system, and WaveNet \cite{oord2016wavenet} demonstrated one way to directly synthesise a waveform, although optimising the DNN loss function at the waveform sample level would seem to be only very loosely related to perceived error.



Burying all of the speech signal processing inside a statistical model might not be desirable for some applications. Traditional representations used in vocoders, such as source and filter, are intuitive and amenable to manipulation. Pitch, for example, can be easily tuned according to listener preference by applying simple scaling to F0. The spectral envelope can be frequency-warped in order modify speaker identity. Duration can be scaled to manipulate speaking rate. Such techniques provide very simple and efficient ways to generate a variety of speakers and styles from a single statistical model.







\section{Attributing degradations to modelling assumptions by performing selective comparisons}
\label{ch18-diagnose_method}


The previous section discussed why vocoded speech (analysis-synthesis) is worse than natural speech. But degradations in synthetic speech -- that is, the ways in which synthetic speech is worse than natural speech -- are not limited to the vocoder.  In general, speech generated from text using a statistical parametric model is more degraded than vocoded speech in terms of signal quality, expressivity, similarity to the original speaker, and intelligibility. Model-generated prosody can be inappropriate or unconvincing. Synthetic speech is generally judged as significantly less natural than vocoded speech, and can be unpleasant to listen to over longer periods \cite{wester2016evaluating}.

The overall quality is a consequence of myriad interacting factors. The remainder of section \ref{ch18-diagnose_method} describes a general methodology that can be used to tease apart the effects of different modelling assumptions. In section \ref{ch18-diagnose_modelling}, we review selected findings from the literature in the light of the described methodology, and ask what they tell us about the effects of common modelling paradigms and assumptions.


\subsection{Basic comparison methodology}
The basic principle for measuring the effects of different speech-synthesis design choices is to contrast the output from two comparable speech synthesis systems; this type of evaluation is widespread.
If multiple aspects of a TTS system are changed incrementally in sequence, a chain of different synthesisers is obtained, and the relative severity of the different assumptions and simplifications involved can be studied. 

In practice, all evaluations are influenced by the context in which they take place:
\begin{enumerate}
\item The \textbf{training data} used, discussed in section \ref{ch18-diagnose_data}
\item The \textbf{evaluation methodology}, discussed in section \ref{ch18-diagnose_evaluation}
\item The \textbf{surrounding model}, discussed in section \ref{ch18-diagnose_modelcontext}
\item The \textbf{output generation method}, discussed in section \ref{ch18-diagnose_generation}
\end{enumerate}
Effects of mathematical modelling assumptions, making up the bulk of the TTS design aspects to be discussed, are covered in section \ref{ch18-diagnose_modelling}.

\subsection{The data}
\label{ch18-diagnose_data}
Any given model or approach does not necessarily work equally well for all datasets. Apart from well-known (but not well-understood) variations caused by speaker characteristics, the amount of data obviously has an effect. For a small dataset, a simple decision tree may outperform a more complex neural network, but the complex model may ultimately give best performance once there is enough data. In practice, most investigations involve just a single speaker and a single fixed-size dataset, which limits the generality of their conclusions.

The quality of the data will also have an affect: transcription errors, signal issues including recording noise, reverberation, compression or transmission artefacts, or even problems with the speech articulation (e.g., disordered and dysarthric speech). This is a broad topic which has had limited systematic exploration \cite{yamagishi2008robustness,karhila2014noise,bollepalli2013effect,creer2013building}.

In general, current commercial practice prevails even in much academic research: fairly large quantities of specially-designed, cleanly recorded, and well-transcribed material is used, even when this leads to a bland and sometimes unnatural speaking style.

\subsection{The evaluation}
\label{ch18-diagnose_evaluation}
Objective comparisons rely on analytic criteria that can be computed and perhaps even optimised automatically, while subjective comparisons generally take the form of listening tests that require careful setup and a group of human evaluators comparing or rating speech stimuli.

Listening tests can produce categorical or numerical outcomes. The former is most common for preference tests or difference detection (discrimination) tests, which tend to be the most sensitive in detecting differences. Tests that directly produce numerical performance measures include mean opinion scores (MOS) \cite{itu1996telephone} and MUSHRA \cite{itu2015method}, of which the latter has been found to be more sensitive \cite{ribeiro2015perceptual}. These paradigms are likely be a better choice for assessing the relative severity of different choices in a spectrum of models by measuring effect sizes, rather than only identifying statistically significant differences. It is possible to infer similar information from categorical judgements, such the ratios of similar vs.\ different judgements between pairs of tested systems. These can be interpreted as relative system similarity, and thus give a picture of listeners' perceptual space. By applying \emph{multidimensional scaling} (MDS) \cite{borg2005modern} to similar-different judgements, systems under test can be located in a continuous-valued perceptual space \cite{merritt2015attributing,merritt2013investigating,henter2014measuring}.

In any evaluation, one must carefully consider the task, which includes the question that listeners are asked to answer, and how the results are analysed \cite{wester2015using}. It is seldom effective to ask listeners to attend to very specific aspects of speech; see also \cite{merritt2016overcoming}. Instead, most evaluations use broad and non-specific formulations such as ``Rate the quality of the following speech samples''. Given that nearly all relevant experimental results are based on this type of question, this chapter is restricted to considering the effects of different design choices on generic ``quality'' or ``naturalness''. That said, the question asked in a subjective test \emph{is} important and can influence the outcome, all other factors being equal, cf.\ \cite{dall2014rating}. 
If we were to consider limitations in, e.g., synthetic speech intelligibility instead of naturalness, we would find that system rankings change \cite{king2014measuring}. This shows that design choices can impact different metrics in quite distinct ways.

Objective comparisons have obvious advantages of being fast, cheap, and straightforward, but it is notoriously difficult to devise objective criteria that correlate adequately with human judgements \cite{hinterleitner2017quality}. Consider the case of the global variance (GV) of speech parameter trajectories. Na\"{\i}ve synthesis systems typically generate parameter trajectories with much smaller dynamic range -- that is, less global variance -- than those of natural speech, and the result is perceived as ``muffled''. Reduced variance is commonly taken as evidence of a loosely-defined issue known as \emph{over-smoothing}. By changing either the model or the generation procedure \cite{zen2007hmm,shannon2013fast}\cite[Sec.\ 6.4.4]{shannon2014probabilistic} to match the global variance observed in training data, perceptual output quality is significantly improved. This has been replicated numerous times, with different synthesisers, datasets, and different techniques for re-instating the variance \cite{toda2007speech,silen2012ways,toda2012implementation,nose2016efficient}. However, acoustic model likelihood will actually be reduced by these techniques: an objectively inferior model produces perceptually superior output.


Learning an objective measure from existing human judgements is a more promising direction. On unseen data, the best current machine-learning predictors have a Spearman's correlation coefficient around 0.6 between predicted and actual per-stimulus mean opinion scores from previous Blizzard Challenge evaluations \cite{yoshimura2016hierarchical}.

Even the best objective criteria frequently fail to identify meaningful differences between stimuli, instead predicting that most stimuli will be judged as close to the average performance, even though the mean scores assigned by human listeners are distributed over a wide range from bad to good. For this reason, objective criteria such as mel-cepstral distortion or parameter estimation objective function value (e.g., data likelihood) should be reserved for when a very large number of comparisons have to be made, such as when tuning parameters during system development \cite{kominek2008synthesizer}. When we are interested in what actually works in practice -- as in this chapter -- there is no substitute for carefully elicited human judgements.

\subsection{The surrounding model}
\label{ch18-diagnose_modelcontext}
The accuracy and properties of the surrounding model will affect which issues are audible and identifiable. The standard approach is to start from a low-accuracy model -- most commonly, a complete baseline text-to-speech system -- in which case the output is a lower bound on the maximum performance achievable: ``By construction, we know that it is possible to do at least this well''. The study \cite{watts2016hmms} is a prime example of this approach, and will be cited extensively in this chapter. 

A less common approach, but one that can provide insights into aspects of performance that are not audible in the standard approach (perhaps because they are obscured by other degradations), is to start from a high-accuracy speech model. Since we do not actually have such a model available, the effects of different assumptions and design choices can only be simulated by manipulating parameters extracted from natural speech recordings; these samples are assumed to be random samples from the true (but unavailable) model. \cite{henter2014measuring} is the most prominent example of this type of study, in terms of the results discussed in this chapter.

It should be emphasized that conclusions from manipulation-based studies generally only apply in the limit of highly accurate models, and it might be possible for less accurate models to surpass them. As an example, we have seen in section \ref{ch18-diagnose_evaluation} above that low-accuracy acoustic models designed to inflate the global variance are often perceptually superior to models that maximise the likelihood of the training data. Manipulation-based approaches might also introduce unintended processing artefacts. In the study \cite{henter2014measuring}, several control conditions were introduced, and the listener scores given to these systems showed that vocoding and duration manipulation on their own could not explain the performance degradations uncovered.

Some investigations consider a wide spectrum of different systems and evaluate both modelled and modified speech stimuli together, including some that are above the typical analysis-synthesis top line. An example that typifies this approach is \cite{merritt2015attributing}. Hybrid evaluation approaches also exist: when evaluating acoustic models, it is not uncommon to generate speech parameters based on a highly-accurate duration model, namely oracle durations copied from held-out natural speech, as in \cite{watts2016hmms} amongst many others. It is thus not possible to interpret the outcome from such evaluations as either lower or upper bounds, or as an indication of what could be expected for an end application. This caveat is especially relevant for prosody, where durations can have a substantial impact on perception.

\subsection{The output generation method}
\label{ch18-diagnose_generation}
Finally, the manner in which the model generates speech parameters can also have a very substantial impact on the perceived properties of the output speech. 
In principle, natural speech can be seen as samples from an unknown and highly complex ``true'' statistical model of speech. It may therefore seem compelling to generate speech by sampling from trained models. Unfortunately, this exposes severe issues with most parametric synthesis models, and randomly generated output sounds notoriously poor: speech sound durations are highly idiosyncratic, while acoustics change so randomly and rapidly that the output sounds warbly or bubbly. Only with the very recent WaveNet -- which is waveform-level rather than statistical parametric speech synthesis -- have sampling-based methods been able to generate good signal quality \cite{oord2016wavenet}.

In order to avoid the issues associated with random sampling from poor models, virtually every practical TTS system instead uses a deterministic output generation criterion that returns a carefully curated, identical output each time. For acoustic models, the most widespread criterion is so-called \emph{most likely parameter generation} (MLPG) \cite{tokuda2000speech}.\footnote{MLPG is sometimes read as ``maximum likelihood parameter generation'', but this is something of a misnomer, since a likelihood denotes the probability assigned to a fixed dataset as the model changes, not the probability of variable data for a fixed model as considered here \cite{ling2012minimum}; the same generation principle has also been called ``maximum output probability parameter generation'' (MOPPG) and ``standard parameter generation'' \cite{henter2014measuring,ling2012minimum,shannon2014probabilistic}.}
This is based on synthesising output from the most probable output sequence under the current speech model, given the input text.

In practice, the most probable output (or \emph{mode}) is very difficult to estimate from real data, and can be slow or infeasible to compute even for a fitted statistical model. Assumptions are used to circumvent this -- for instance, instead of integrating over all possible utterance durations and all possible paths through the hidden state space in an HMM, only a single path through the state space is used. To what degree this choice degrades or improves the output has not been studied in depth. Conveniently, whenever the distribution over output trajectories is Gaussian (which is frequently the case with conventional models and single paths through the state-space during generation), the most likely trajectory is simply the mean of this Gaussian model, which can be computed using the algorithms in \cite{tokuda2000speech}. It is well established that predicting and synthesising from the resulting mean trajectory sounds much better than random sampling \cite[for example]{uria2015modelling}. In addition, since estimating means is statistically straightforward, it is possible to estimate the mean trajectory of highly accurate models of speech as well; this was done in the study by \cite{henter2014measuring}, with the finding that the mean of highly accurate models is perceptually inferior to random samples from the same model, at least in the domain of speech parameters derived from STRAIGHT. This shows that there is a very substantial interaction between the modelling assumptions and generation techniques: for poor models, random examples sound worse than the mean, while for accurate models the reverse is true.


In practice, there are other deterministic generation schemes that tend to be subjectively preferred over the raw (approximate) most probable parameter trajectory of a maximum-likelihood fitted acoustic model. These revised output generation procedures are generally based on the idea of compensating for the lower-than-expected global variance (GV) of speech produced by the standard deterministic procedure.

The GV deficiency arises because, somewhat surprisingly, the ``most probable'' output sequence is not guaranteed to be a typical example of speech. It may actually be far from natural.
As stated in section \ref{ch18-diagnose_modelcontext} above, insufficient GV in synthesised speech appears to correlate with poor subjective scores, and the GV can be boosted either by changing the model to yield more appropriate GV under MLPG, or by changing the generation principle instead. The latter usually involves changing the MLPG output prior to playout, which is frequently called \emph{post-processing}, or \emph{postfiltering} after an early post-processing paradigm in speech compression \cite{ramamoorthy1984enhancement}. (Note, however, that correlation does not imply causation, and issues with generation methods being GV-deficient may not be the only reason that GV-boosted output frequently is considered perceptually superior.) Some other approaches to this are variance scaling \cite{silen2012ways} and methods incorporating global affine transformations \cite{toda2012implementation,nose2016efficient,ling2012minimum}.

As a final note, the generation methods used also affect what objective evaluation criteria that are seen as most appropriate: whereas maximum likelihood, or similar criteria which assess the accuracy of the predicted speech parameter distribution, are most informative for the case of sampling-based generation, methods that consider only the generated output (such as RMSE or MCD) might be superior for deterministic techniques. Since post-processing often has an adverse effect on standard objective metrics, it is common to apply post-processing only for subjective listening tests, but not when making objective comparisons.





\section{Shortcomings of Statistical Models for Speech Synthesis}
\label{ch18-diagnose_modelling}


Having described in section \ref{ch18-diagnose_method} above a general methodology for attributing degradations to modelling assumptions and other design choices, and for assessing their severity, this section will present empirical findings regarding:
\begin{enumerate}
\item Duration modelling, in section \ref{ch18-diagnose_duration}
\item Machine-learning paradigm (acoustic regression model), in section \ref{ch18-diagnose_ml}
\item Across-context averaging, in section \ref{ch18-diagnose_crosscontext}
\item Distribution and dependence assumptions, in section \ref{ch18-diagnose_dependence}
\item Joint or separate stream modelling, in section \ref{ch18-diagnose_streams}
\item Temporal modelling, in section \ref{ch18-diagnose_temporal}
\item Optionality, in section \ref{ch18-diagnose_optionality}
\end{enumerate}
We begin by briefly introducing the two main comparative studies upon which the subsequent discussion is based.

\subsection{Key comparative studies}
\label{ch18-diagnose_studies}
\cite{watts2016hmms} and \cite{henter2014measuring} are both side-by-side MUSHRA tests that analyse the relative impacts of a chain of different modelling assumptions and design choices. The former focussed on modelled speech (a low-naturalness operating point) and the latter on manipulated speech (a high-naturalness operating point).

The study in \cite{watts2016hmms} compared a number of different text-to-speech systems designed to interpolate between, atone end, a state-of-the-art decision-tree-based speech synthesis system (HTS, \cite{zen2007hmm}) and, at the other end, a recent deep-neural-network-based synthesiser (Merlin, \cite{wu2016merlin}) using feedforward DNNs. The systems were all trained on the same database and used the same vocoder (STRAIGHT). All used oracle durations from held-out natural speech, making these studies an investigation into the performance of different acoustic modelling techniques.

The main aim of \cite{watts2016hmms} was to identify the key factors that contribute to the empirically-observed improvement in performance of newer DNN-based TTS systems over established decision-tree-based synthesisers like HTS. The implementational differences between systems from the two paradigms are not limited to the machine-learning technique, but include many additional choices; those examined were:
\begin{enumerate}
\item Regression model: decision trees (DT) or feedforward deep neural networks (NN) (section \ref{ch18-diagnose_ml}).
\item Temporal granularity: piece-wise constant for each sub-phone state, or changing every frame (section \ref{ch18-diagnose_temporal}).
\item Stream modelling: each stream of speech parameters can be predicted by a separate regression model or all can be predicted together by a single, joint model (section \ref{ch18-diagnose_streams}).
\item Variance model: the variance used during generation can be predicted by the regression model or can be a global constant (section \ref{ch18-diagnose_dependence}).
\item Duration-dependent input features: whether duration is used by the acoustic regression model (section \ref{ch18-diagnose_duration}).
\item Trajectory enhancement method: global variance modelling (GV) \cite{toda2007speech} or regular mel-cepstral domain postfiltering for formant enhancement (PF) \cite{yoshimura2005incorporating}. (sections \ref{ch18-diagnose_generation} and \ref{ch18-diagnose_crosscontext}). 
\end{enumerate}

The configurations of the different TTS systems built, and specifically their differences in terms of the above-mentioned aspects, are listed in Table \ref{ch18-diagnose_hmms2dnnssysts}. 
\begin{table}[!t]
\centering
\begin{tabular}{|l|l|l|l|l|l|l|}
\hline 
\textbf{ID} & \textbf{Model} & \textbf{Resolution} & \textbf{Streams} & \textbf{Variance} & \textbf{Dur.\ dep.} & \textbf{Enhancement}\tabularnewline
\hline 
\hline 
V & - & - & - & - & - & -\tabularnewline
\hline 
D1 & DT & state & separate & local & no & GV\tabularnewline
\hline 
D2 & DT & state & separate & local & no & PF\tabularnewline
\hline 
N1 & NN & state & separate & local & no & PF\tabularnewline
\hline 
N2 & NN & state & separate & global & no & PF\tabularnewline
\hline 
N3 & NN & state & joint & global & no & PF\tabularnewline
\hline 
N4 & NN & frame & separate & global & no & PF\tabularnewline
\hline 
N5 & NN & frame & joint & global & no & PF\tabularnewline
\hline 
N6 & NN & frame & joint & global & yes & PF\tabularnewline
\hline 
\end{tabular}
\caption[An overview of the TTS systems compared in \protect\cite{watts2016hmms}.]{An overview of the TTS systems compared in \protect\cite{watts2016hmms}, showing their IDs and the factors that were successively altered in order to step from HTS (``D1'') to Merlin (``N6''). More information about the different factors is provided in the text and in \protect\cite{watts2016hmms}. ``V'' is analysis-synthesised speech acting as a top line reference in the listening test.}
\label{ch18-diagnose_hmms2dnnssysts}
\end{table}

\begin{figure}[!t]
\centering
\includegraphics[width=\columnwidth]{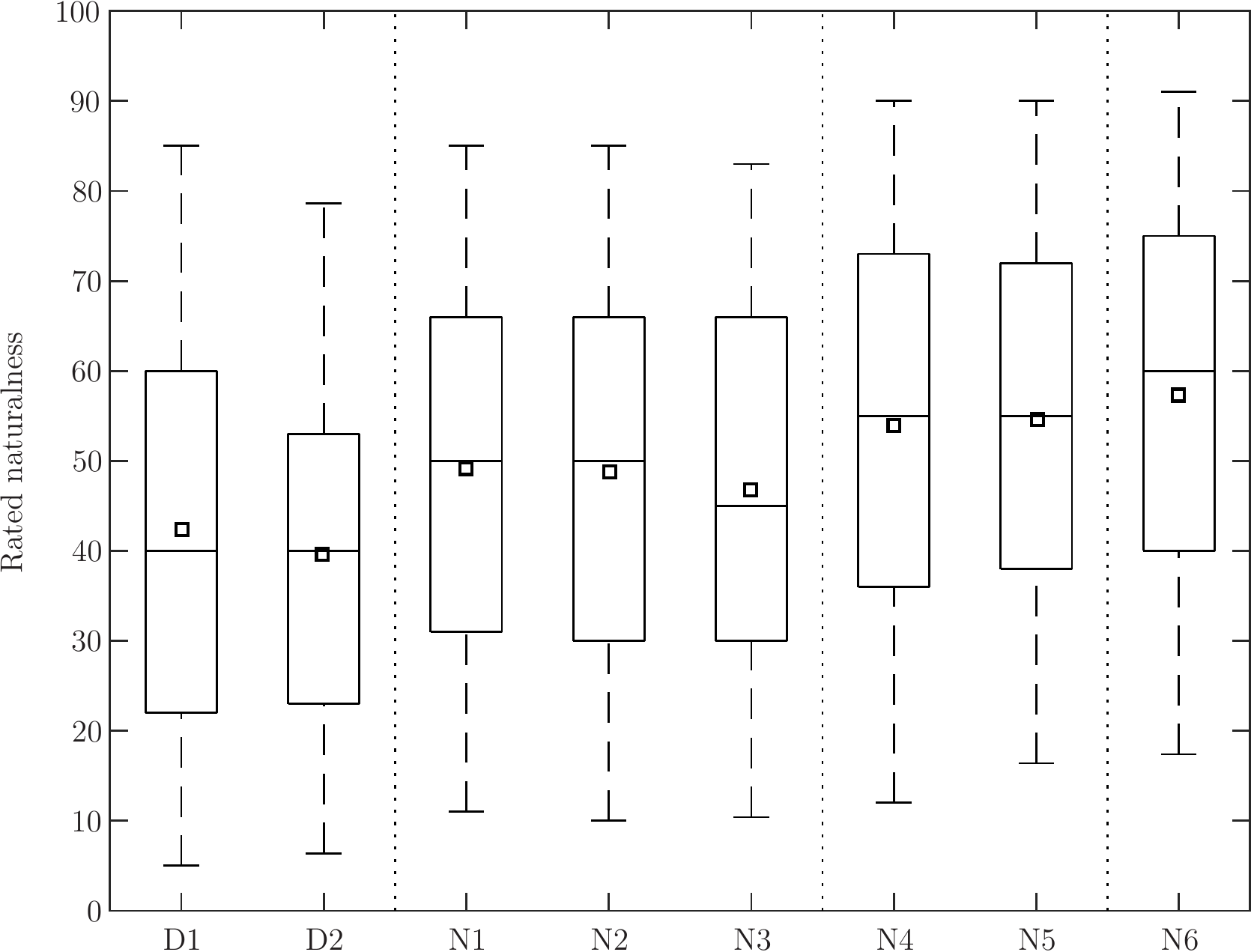}
\caption[Box plot of aggregate listener naturalness ratings from the MUSHRA test in \protect\cite{watts2016hmms}.]{Box plot of aggregate listener naturalness ratings from the MUSHRA test in \protect\cite{watts2016hmms}. Condition labels are as in table \protect\ref{ch18-diagnose_hmms2dnnssysts}; ``V'' is omitted as it was rated at 100 (maximally natural) more than 95{\%} of the time. For the boxes, middle lines show medians, box edges are at quartiles, while whiskers extend to cover all but 5{\%} of data on either side. Squares denote the mean rating. Dotted lines separate systems into sets, where systems within a set exhibited no statistically significant differences in rating, while all cross-set differences were statistically significant.}
\label{ch18-diagnose_hmms2dnnsmushra}
\end{figure}

After building eight different text-to-speech systems as listed in Table \ref{ch18-diagnose_hmms2dnnssysts}, output from these systems plus vocoded natural speech were compared in a MUSHRA test. In the test, 20 native, paid listeners each scored parallel system output on 20 phonetically-balanced sentences (selected for each participant in a balanced manner from a pool of 70), for a total of 400 parallel ratings. The results of the test are illustrated in figure \ref{ch18-diagnose_hmms2dnnsmushra}. The stimuli and ratings are freely and permanently available online.\footnote{\url{doi:10.7488/ds/1316}} Upon applying double-sided Wilcoxon signed-rank tests to all system pairs, with a Holm-Bonferroni correction \cite{holm1979simple} to keep the familywise error rate below $\alpha=0.05$, the different systems studied separated into five distinct sets, such that all between-set comparisons were statistically significant, whereas all within-set comparisons were not.
The sets are delimited by dotted vertical lines in figure \ref{ch18-diagnose_hmms2dnnsmushra}. From these results, it seemed that the major gains in synthesis performance between systems ``D1'' and ``N6'' coincided with the switch from decision trees to neural networks in the regression model and with the change from state-level to frame-level time granularity. Adding duration-derived features also made a significant difference, although the effect size was smaller; it is unclear if this would apply when using predicted rather than oracle durations. While GV is generally considered superior to regular formant-enhancement postfiltering, that difference was by comparison not so large as to be significant in this investigation, though the difference between ``D1'' and ``D2'' was judged as significant if a per-subject score normalisation was introduced \cite{watts2016hmms}. The implications of the study findings are discussed more in-depth in section \ref{ch18-diagnose_ml} onwards.

The second main study \cite{henter2014measuring} evaluated only natural and manipulated speech, making it an investigation of the upper limits placed on naturalness by a number of modelling assumptions and design choices in speech synthesis, in the context of a highly accurate model. The central innovation was to use a carefully purpose-recorded database of repeated speech, called the Repeated Harvard Sentence Prompts (REHASP) corpus version 0.5, where the same sentence prompt was read aloud multiple times by the same speaker in identical conditions. Each recording can then be seen as a statistically independent sample from the same true speech distribution for that particular sentence. This database is freely available.\footnote{\url{doi:10.7488/ds/39}}

\begin{figure}[th]
\centering
\hfill
\subfigure[Aligned repetitions\label{ch18-diagnose_manipulationsa}]{\qquad\includegraphics[width=0.15\columnwidth]{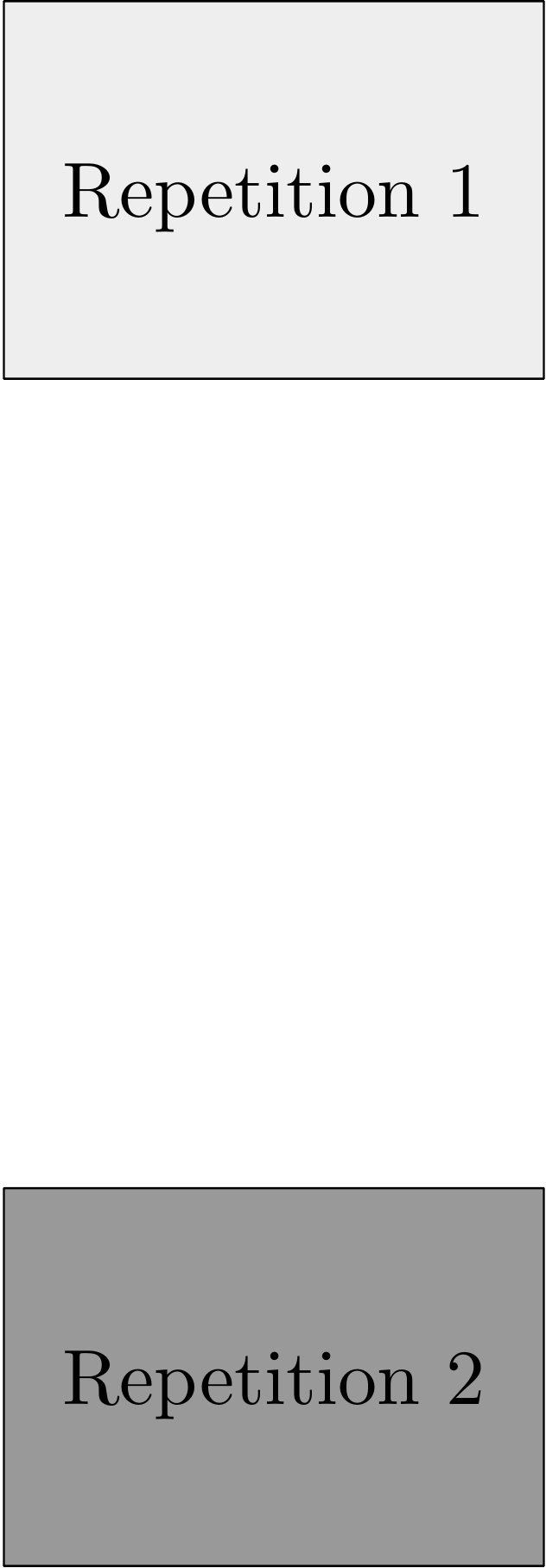}\qquad}
\hfill
\hfill
\subfigure[Mean speech\label{ch18-diagnose_manipulationsb}]{\qquad\includegraphics[width=0.15\columnwidth]{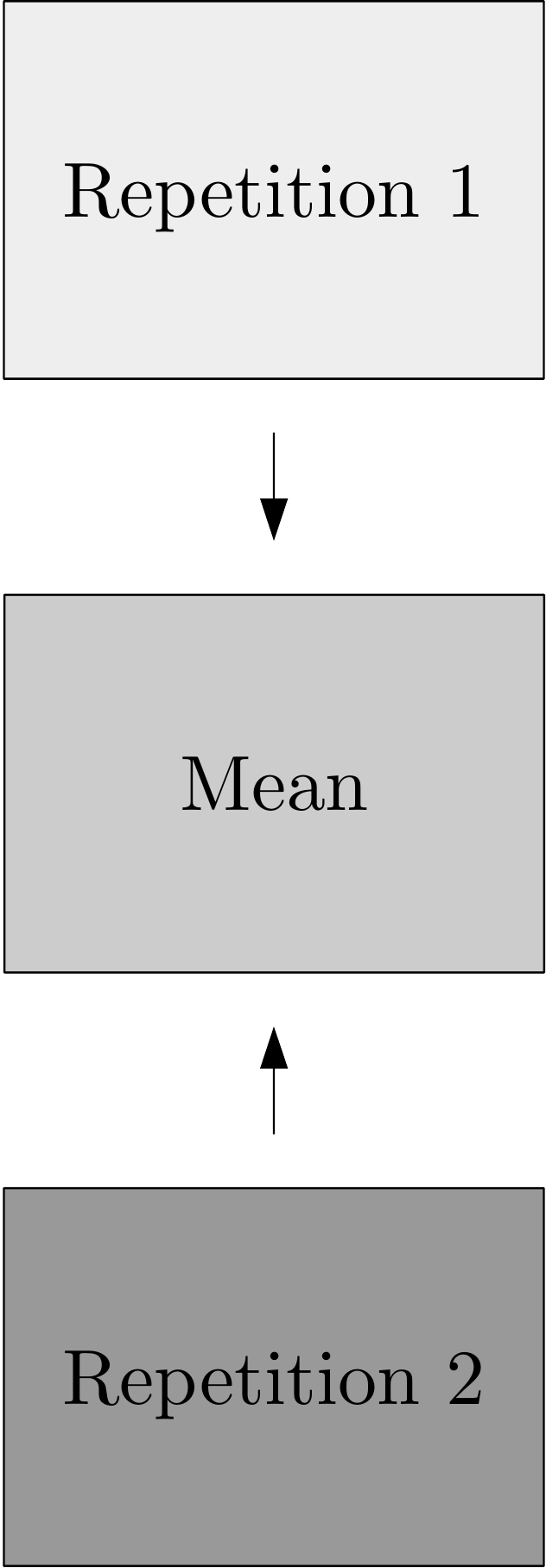}\qquad}
\hfill
\hfill
\subfigure[Chimeric speech\label{ch18-diagnose_manipulationsc}]{\qquad\includegraphics[width=0.15\columnwidth]{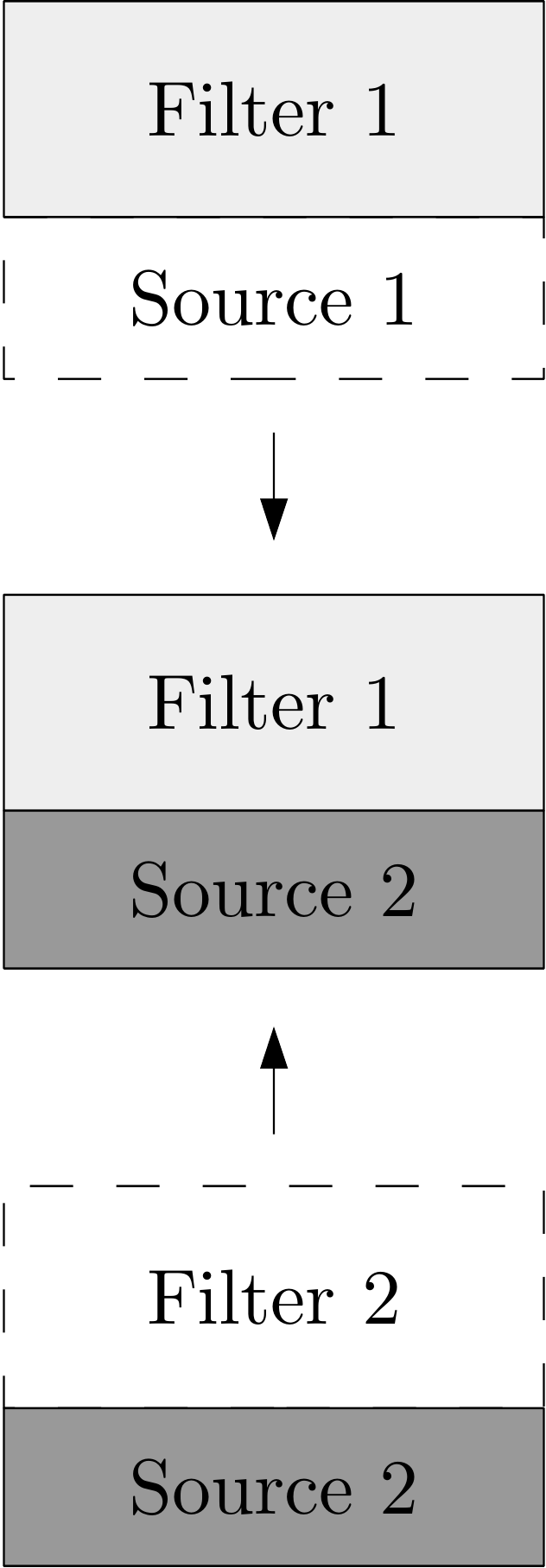}\qquad}
\hfill
\caption[Manipulating repeated speech.]{Manipulating repeated speech. After obtaining time-aligned independent readings (repetitions) of the same prompt, shown in \ref{ch18-diagnose_manipulationsa}, the aligned speech-parameter matrices can be blended into an estimate of the true average speech, as in \ref{ch18-diagnose_manipulationsb}, or stitched together to form chimeric speech that represents randomly-sampled speech output from a highly accurate model with certain independence assumptions, such as conditional source-filter independence in \ref{ch18-diagnose_manipulationsc}.}
\label{ch18-diagnose_manipulations}
\end{figure}

By applying dynamic time warping, all repetitions of the same prompt were also made to have the same timings. (The mathematical interpretation of this is that, for each frame, all time-warped repetitions were in the same state in a left-right state-space model at that frame. Due to how HMMs are defined, different parameter trajectories can then be treated as conditionally independent.) This allowed the different repetitions to be combined in various ways, either creating \emph{chimeric speech} by taking the trajectories of different features from different, independent repetitions, or blending all repetitions into one grand average; see figure \ref{ch18-diagnose_manipulations}. The result was manipulated speech stimuli approximating the output of highly accurate trajectory models, under various conditional-independence assumptions (no independence versus independence between source and filter, parameter streams, and filter coefficients) and generation methods (random sampling or taking the mean). Table \ref{ch18-diagnose_rehaspconds} details how the most relevant conditions from the study were constructed, as well as their interpretation.
\begin{table}[!t]
\tabcolsep=0.12cm 
\centering
\begin{tabular}{|l|l|c|c|c|c|l|l|l|}
\hline 
\multicolumn{2}{|c|}{\textbf{Condition}} & \multicolumn{4}{c|}{\textbf{Parameter source repetitions}} & \multicolumn{3}{c|}{\textbf{Interpretation}}\tabularnewline
\hline 
\textbf{ID} & \textbf{Description} & \textbf{Dur.} & \textbf{LF0} & \textbf{BAPs} & \textbf{MCEPs} & \textbf{Domain} & \textbf{Model} & \textbf{Generation}\tabularnewline
\hline 
\hline 
N & Natural waveform & - & - & - & - & Waveform & True & Sampling\tabularnewline
\hline 
V & Vocoded & a & a & a & a & Param. & True & Sampling\tabularnewline
\hline 
D & Time-warped & b & a & a & a & Param. & $\mathcal{M}_{\text{D}}$ & Sampling\tabularnewline
\hline 
SF & Source/filter indep. & b & a & a & c & Param. & $\mathcal{M}_{\text{SF}}$ & Sampling\tabularnewline
\hline 
SI & All streams indep. & b & a & d & c & Param. & $\mathcal{M}_{\text{SI}}$ & Sampling\tabularnewline
\hline 
I & MCEPs indep. & b & a & a & $\ast$ & Param. & $\mathcal{M}_{\text{I}}$ & Sampling\tabularnewline
\hline 
M & MCEPs averaged & b & a & a & mean & Param. & $\mathcal{M}_{\text{D}}$\textendash{}$\mathcal{M}_{\text{I}}$ & Mean\tabularnewline
\hline 
\end{tabular}
\caption[An overview of the main speech manipulations investigated in \protect\cite{henter2014measuring}.]{An overview of the main conditions (speech manipulations) investigated in \protect\cite{henter2014measuring}, showing their ID and description, where different parameter trajectories were sourced for each manipulation and how they are to be interpreted. For the parameter trajectory sources, each different letter represents a different source repetition that was used; ``$\ast$'' means that each coefficient trajectory was taken from a different repetition in the database, while ``mean'' is an average over all repetitions. The interpretation-related fields distinguish the domain of the synthesis (waveform or speech parameters), the statistical model used and the output generation method.}
\label{ch18-diagnose_rehaspconds}
\end{table}

\begin{figure}[!t]
\centering
\includegraphics[width=\columnwidth]{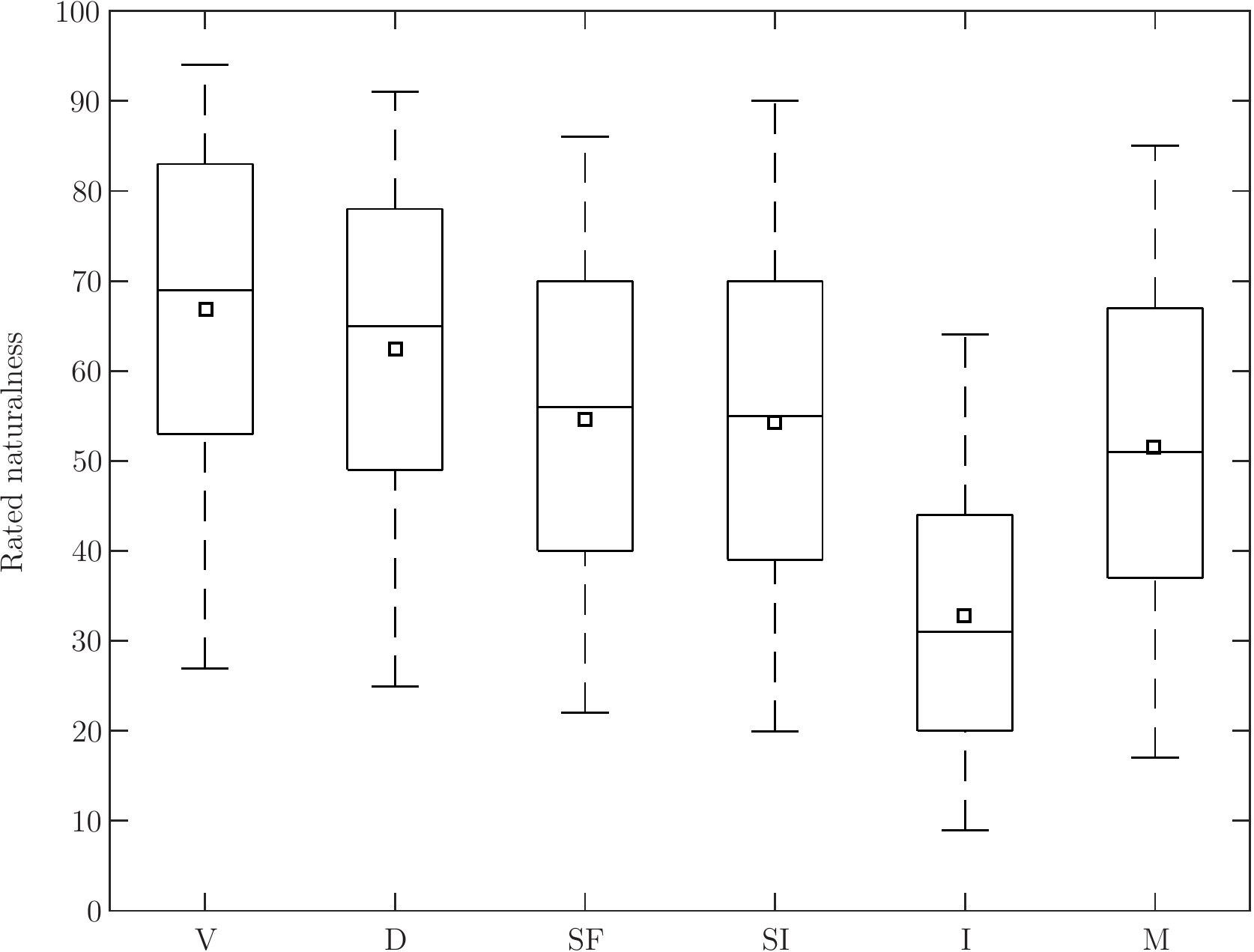}
\caption[Box plot of aggregate listener naturalness ratings from the MUSHRA test in \protect\cite{henter2014measuring}.]{Box plot of aggregate listener naturalness ratings from the MUSHRA test in \protect\cite{henter2014measuring}. Condition labels are as in table \protect\ref{ch18-diagnose_rehaspconds}; ``N'' is omitted as it always was rated at 100 (completely natural). For the boxes, middle lines show medians, box edges are at quartiles, while whiskers extend to cover all but 5{\%} of data on either side. Squares denote the mean rating. All system pairs showed significant differences in rating except (SF, SI) and (SI, M).}
\label{ch18-diagnose_rehaspmushra}
\end{figure}

30 native, paid listeners compared the naturalness of different manipulated speech examples for the same prompt in a MUSHRA test. The results of this test, based on a total of 549 parallel ratings, are summarised in figure \ref{ch18-diagnose_rehaspmushra}. Bonferroni-corrected pairwise $t$-tests found all system pairs to be significantly different at the 0.01 level, except (SF, SI) and (SI, M). The results thus lend some insight into how much the tested assumptions might limit speech synthesiser naturalness, as discussed in the remainder of this chapter. It should, however, be pointed out that the results, strictly speaking, only are known to be valid for the specific speech parameterisation used in the study, which was based on legacy STRAIGHT with a mel-cepstrum representation of the STRAIGHT spectrogram.

\subsection{Duration modelling}
\label{ch18-diagnose_duration}
A probabilistic model of speech parameter sequences is usually factored into a duration model and an acoustic model, where durations typically have to be predicted first during synthesis. We will therefore start by briefly discussing the duration model, before acoustic modelling.

While duration is an important component of synthesising speech there have been relatively few studies that have tried to isolate the effects of different design choices in modelling duration, compared to the extensive literature on acoustic modelling.
The simplest possible duration model is to assume that all (sub-state) durations are independent and given by a discrete, no-skip, left-right Markov chain. This was a common model in early decision-tree-based synthesis. In this model, advancing to the next state is essentially decided by a coin toss (a Bernoulli random variable), so that individual state durations implicitly follow a geometric distribution. This model is a poor fit with actual durations in state alignments on training data (for one thing, the most probable duration is always a single frame), but is acceptable for synthesis as long as the expected duration is used at generation time, noting that HMM training (maximum likelihood parameter estimation) of duration parameters in this case reduces to matching the expected duration of the model with the mean durations observed in the training data.

\cite{zen2007hidden} proposed improving the duration model by replacing the Markov model over the state space (and the implicit geometric duration distributions it entails) with a semi-Markov model, which is memoryless given the current state \emph{and} how many time steps the process has spent in the current state. This allows state durations to follow any discrete-valued stochastic distribution. The resulting construction is known as a \emph{hidden semi-Markov model} (HSMM). In practice, it is often assumed that durations follow a Gaussian distribution, even though this distribution is continuous-valued (not discrete) and can attain negative values. This choice was seen to improve the subjective quality of synthesised speech in the small listening test in \cite{zen2007hidden}, and is incorporated as standard in HTS, though the relative importance of the improvement has not been well studied in relation to other problem sources. HSMMs were also recently used for EM-based realignment in a feedforward DNN-based speech synthesiser \cite{tokuda2016temporal}. Either way, since the change does not directly alter the fundamentals of how durations are predicted -- predictions are still based on the mean duration of the state duration, which still is estimated as the sample mean over aligned phones allocated to the relevant decision tree node -- any improvements an HSMM brings over an HMM are likely to be due to improved alignments indirectly benefiting both duration models and acoustic models, by more accurately associating frames with a suitable state in a contextual phone.

A substantially different approach to speech synthesis duration modelling was recently suggested by \cite{ronanki2016median}. They replaced the parametric Gaussian distributions of traditional HSMMs by a non-parametric distribution predicted by the same recurrent neural network that also predicted acoustic parameters, and used distribution quantiles (rather than the mean) for output generation. It is, however, too early to tell if this approach will bring subjective improvements in the long run.

\subsection{The machine learning paradigm}
\label{ch18-diagnose_ml}
The machine learning paradigm used to predict speech properties from text-extracted linguistic features is the other major factor that affects duration modelling. This area has seen a major paradigm shift in recent years, with decision trees being replaced by deep and recurrent learning techniques. For acoustic modelling, specifically, it has generally been found \cite{wang2016comparative} that decision-tree regression is inferior to deep feedforward neural networks \cite{zen2013deep,zen2014deep,hashimoto2015effect,watts2016hmms}, which in turn tend to be outperformed by recurrent neural network techniques \cite{fan2014tts,zen2015unidirectional}, of which the long short-term memory \cite{hochreiter1997long} has received the most attention. The study by \cite{watts2016hmms} is especially elucidating, since it compared the relative impact of many different aspects that distinguish conventional decision-tree-based systems from recent feedforward-DNN-based TTS, and found that the change in machine learning paradigm yielded one of the most notable improvements (both in numerical size and statistical significance) observed in their quantitative MUSHRA listening test. 

While feedforward or recurrent neural networks have become central to high-quality speech synthesisers in many applications, there are alternative regression techniques as well, such as random forests \cite{black2015random}. They were also able to show improved modelling accuracy over only using a single decision tree for regression.


\subsection{Cross-context averaging}
\label{ch18-diagnose_crosscontext}
This issue arises when the synthesis approach fails to distinguish frames that, conditioned on the input text, should be realised as acoustically distinct, thus treating training examples as interchangeable when they are not. Cross-context averaging has been found to be particularly harmful to synthesis quality \cite{merritt2015attributing,merritt2016overcoming}. It is easy to see that this situation can occur in decision-tree-based approaches, which rely on clustering training data frames together and assigning them to leaves in a tree: only aggregate properties of these clusters are used in synthesis.

Cross-context averaging is not unique to decision trees, however. A similar effect can arise with any machine learning paradigm, in cases where the available input features (linguistic or otherwise) fail to distinguish speech contexts that should be given acoustically distinct realisations, for instance by not properly separating between stressed and unstressed instances of the same word.
Mathematically, if there is structured variation that cannot be predicted due to conflating and averaging speech across contexts, this variation will instead be absorbed by the model: the noise term in a Gaussian model. Since the most likely output is the mean (which ignores the noise term) cross-context averaging is likely contribute to the deficient GV of generated parameter trajectories. At the same time the over-inflated noise term is also likely to contribute to the unappealing, noisy and unstructured behaviour of random sampling from such a model.


It was observed that scaling up the variance of synthesised speech to match the training data GV did not completely undo the detrimental effects of the averaging \cite{merritt2015attributing}, presumably because the model is incapable of recreating the missing contextual distinctions. However, there are methods to mitigate cross-context averaging, often referred to as \emph{rich-context models}. Several studies support the conclusion that rich context modelling leads to perceptually superior synthesised speech, both in conventional decision-tree-based speech synthesis \cite{yan2009rich,merritt2015deep} and in hybrid synthesis approaches \cite{merritt2016deep}.

A typical rich-context model will distinguish all possible quinphones centred on the current phone. This is not to say that contextual information outside this window cannot be informative as well. Wu et al.\ \cite{wu2016improving,wu2015deep} augmented the standard linguistic features with bottleneck-DNN-derived features that summarise the acoustically most salient information contained in the input features of surrounding frames. Synthesis with the augmented features of 23 surrounding frames produced synthetic speech that was subjectively preferred. It is likely that improvements provided by recurrent neural networks \cite{fan2014tts,zen2015unidirectional} are also due to better propagation of linguistic feature information across frames.

Surprisingly, rich linguistic context alone may never be sufficient for achieving truly convincing speech output, even with utterance-length contextual information, and a model close to true speech. This was explored by \cite{henter2014measuring}, as outlined in section \ref{ch18-diagnose_studies}, performed frame-wise alignment using dynamic time warping of 40 recordings of the same single-sentence prompt, read aloud by a single speaker. By averaging the 40 recordings (frames) for each warped time instant, a stimulus ``M'' was obtained that closely approximates the conditional mean of the ``true'' speech model, given the perfectly-matched utterance-wide linguistic context of the frame. Nevertheless, as seen in figure \ref{ch18-diagnose_rehaspmushra}, the result was significantly inferior in naturalness to both analysis-synthesised speech (``V'') and vocoded speech with time-warped durations (``D''). We can conclude that the operation of averaging -- even when performed over exceptionally comparable acoustic frames whose entire linguistic context is identical -- limits the naturalness that can be achieved.

\subsection{Distribution and dependence assumptions}
\label{ch18-diagnose_dependence}
Most issues discussed so far revolve around how the central tendency (i.e., mean) of the speech is described by the regression model. However, the noise (or covariance) model, used for describing the distribution of deviations from the regression model prediction within each time frame, can also have an impact, particularly for stochastic output generation methods.

The simplest possible description of acoustic frame properties is not to use a model but to minimise a distance measure. In practice, the mean squared prediction error is used almost exclusively. Mathematically, this corresponds to a context-independent isotropic Gaussian with all output dimensions having the same variance. One step up in complexity is a Gaussian with diagonal covariance matrix, so that each dimension has a different (still context-independent, i.e., global) variance. These two setups are commonplace in recent deep learning-based speech synthesis, with the latter being the standard in Merlin \cite{wu2016merlin}. The next step is to allow variances to depend on context, an to predict them with the regression model. This is standard in decision-tree based synthesisers such as HTS \cite{zen2007hmm}. 
In a comparative study \cite{watts2016hmms}, models with global versus context-dependent variance (``N2'' vs.\ ``N1'') were not found to sound very different. 

A further extension is to consider full (i.e., non-diagonal) covariance matrices, though this can easily lead to a model with a very large number of degrees of freedom, susceptible to overfitting. So far, models with full covariance matrices have mostly appeared in decision-tree-based speech synthesisers, for example semi-tied covariance matrices \cite{gales1999semi}.
In any case, the effect on deterministically generated speech output is likely to be subtle, with a minor improvement claimed in \cite{zen2007hmm}. Any improvement comes at
 greatly increased computational complexity of synthesis.

By relaxing the Gaussian assumption, the distribution of speech frame features, and not just their means and variances, can be described and predicted more accurately. In these cases, the most probable output, if used for synthesis, may no longer coincide with the mean of the distribution, 
so the limitations identified for condition ``M'' in \cite{henter2014measuring} may not apply. One example would be real-valued neural autoregressive density-estimators (RNADEs) \cite{uria2014deep}, which let the distribution of feature components depend on the outcome values of preceding components through a neural network. When applied to TTS by \cite{uria2015modelling}, RNADEs were seen to substantially improve held-out data likelihood, and produced deterministically generated speech that was preferred by listeners.

Another non-Gaussian approach to describing speech feature distributions within a time frame is offered by mixture models. In particular, Gaussian mixture models whose parameters are predicted by a deep neural network (an instance of the mixture density networks, or MDNs, described in \cite{bishop1994mdnn}) have given modest improvements \cite{zen2014deep,wang2016gating,henter2016robust}. For computational convenience, these methods usually only consider the most massive mixture component when generating output. The benefit of the remaining mixture components may simply be to absorb difficult-to-explain training data that would otherwise `pollute' the most massive component \cite{henter2016robust}.

If we consider generation based on random sampling, the importance of covariance modelling increases dramatically. \cite{uria2015modelling} uncovered a substantial preference in favour of samples drawn from an RNADE model compared to samples drawn from a similar model with no explicit modelling of conditional dependences between speech features.

\subsection{Separate or joint modelling of parameters}
\label{ch18-diagnose_streams}
Related to dependence assumptions is the topic of between-feature dependence: joint or separate modelling of speech parameter streams, where ``streams'' are subsets of the acoustic features (sometimes also including durations) that are believed to behave similarly in terms of modelling. STRAIGHT features are usually partitioned into three streams, namely (log) F0, the aperiodicity coefficients, and the spectral envelope coefficients (mceps). 
Decision-tree-based synthesis, in particular, can benefit from using separate regression trees for these different feature streams.


In the deep learning paradigm, speech recognition performance has been seen to improve through \emph{multi-task learning}, where the NN is trained to simultaneously solve an additional, related prediction task \cite{qian2015multi} and this idea has also been applied to synthesis \cite{wu2015deep}.

 The conjecture is that lower DNN layers learn to capture more universal and generalisable structure from the data: the additional task acts as a regulariser. So -- unlike decision trees -- it should be beneficial to predict all output streams using a single, joint NN. \cite{chen2015investigation} found that using a single feedforward DNN improved the subjective MOS compared to separate DNNs for each  stream. 
 \cite{watts2016hmms} also considered this distinction, but did not uncover any noticeable differences. 

While separate versus joint stream modelling can influence how cross-stream dependencies are modelled, it is \emph{not} the same as a statistical independence assumption between output features. As a counterexample, most models that predict feature streams using a joint model still assume diagonal-covariance Gaussian distributions. The impact of conditional independence assumptions between streams was investigated by \cite{henter2014measuring}
who found that speech randomly sampled from models that assume parameter streams to be conditionally independent, but otherwise are highly accurate and internally consistent within streams, is subjectively inferior to speech samples from highly-accurate models that also account for cross-stream dependences (conditions ``V'' vs.\ ``SI'' in the study). 

\subsection{Trajectory modelling}
\label{ch18-diagnose_temporal}
Thus far, the discussion of acoustic feature modelling has primarily covered regression techniques, distribution assumptions, and feature dependences within single, isolated time frames. But speech is a stochastic time series, and so it is also necessary to model the temporal evolution of acoustic parameters. This modelling encompasses two parts: assumptions about the stochastic distributions of speech parameter trajectories for a given text, and how the properties of these distributions are made to depend on the text input across time (and in particular the temporal resolution of this dependence).

Many statistical speech synthesis systems account for time-dependence in similar way to automatic speech recognition, by modelling not only so-called `static' frame-wise features, but also the local differences (deltas) and second-order finite differences (delta-deltas) of the frame-wise features: collectively known as \emph{dynamic features}. 
In a Gaussian model of parameter trajectories over time, the use of dynamic features means that the precision matrix must have a band-diagonal structure, and time dependence is restricted to simple, linear correlations.
Statistically, the result is a product-of-experts model, where the most likely output is generated as a compromise between soft constraints on the static and dynamic properties of the trajectory \cite[Sec.\ 3.3.2]{shannon2014probabilistic}. Practically, the consequence is a smoothed output sequence, compared to the piecewise-constant output if only a model of the `statics'.

The relative impact of temporal smoothing on natural speech parameter trajectories was been investigated in \cite{merritt2013investigating} where it was found that the temporal averaging produced by such smoothing had a much smaller effect on naturalness than the cross-context averaging discussed in section \ref{ch18-diagnose_crosscontext} above. This is consistent with \cite{zhang2008improving}, who found overly smooth temporal trajectories less problematic than overly smooth spectra.

A majority of synthesisers ignore the deterministic relationship between static and dynamic feature values during training, only taking them into account during output generation: so, the output is generated from a different model than that created during training \cite{zen2007reformulating,shannon2011effect}. A complementary perspective is that the normalisation constant used during training is incorrect, since it accounts for combinations of static and dynamic feature values that are simply impossible. A consequence of the mismatch is that the trained trajectory model severely underestimates the statistical variation possible in natural speech parameter trajectories, and therefore tends to assign pathologically small probabilities to held-out speech utterances \cite{shannon2011effect}. Using matched and properly normalised models during both training and synthesis may be perceptually superior to the conventional, mismatched approach \cite{zen2007reformulating}.



In decision-tree based models, such as HTS, each context-dependent model comprises a small number of states, and so the statistics of the generated output remain the same for several consecutive frames. Systems exist, however, for which the statistical properties of every frame can be different. A common mechanism is to provide the current position \emph{within} state or phone as an input to the acoustic predictor. This is straightforward, even standard, in neural network-based acoustic models like Merlin, but the same type of feature can with some additional effort also be integrated into synthesisers using Gaussian process regression \cite{koriyama2014statistical} or decision-trees. The canonical example of the latter is Clustergen \cite{black2006clustergen}, where the time resolution is improved by subdividing states (decision tree leaves) by thresholding the position indicator.

For neural networks, it is easy to compare parallel synthesisers that either include or omit these frame-level positional features. The MUSHRA test in \cite{watts2016hmms} found that their inclusion gave one of the most substantial quality improvements, making this a key difference between decision tree approaches such as HTS \cite{zen2007hmm}, and deep learning systems such as Merlin \cite{wu2016merlin}. In contrast, the 10-subject MOS test in \cite{tokuda2016temporal} found very similar performance between a DNN-based synthesiser with frame-level granularity using oracle durations, and a feedforward DNN with state-level granularity and durations predicted from a HSMM duration model, where the DNN-HSMM system had been trained using the generalised EM-algorithm. It is not straightforward to reconcile these two, seemingly conflicting findings. 

Of course, recurrent neural networks, regardless of the input features, are able to learn to model fine-grained positional information, using the internal network state.

\subsection{Optionality}
\label{ch18-diagnose_optionality}
The findings in \cite{henter2014measuring} suggest that, in order to be truly acoustically convincing, speech synthesis needs to move beyond deterministic output generation methods that merely attempt to produce average speech. Many systems already perform post-processing of the generated mean speech to better match the global variance of natural speech, which does increase subjective output naturalness \cite{silen2012ways,toda2012implementation,nose2016efficient}. But, the effects of such processing have not been studied at a high-accuracy operating point, where it seems less likely to be effective.
A more radical change would be to generate speech based on something other than the expected value, for instance by sampling, as discussed in section \ref{ch18-diagnose_generation}, something that we are just starting to see in very recent work.\footnote{For instance, authors' pre-prints not peer-reviewed at the time of writing.}

A third alternative is to allow \emph{optionality} in the speech realisation, which here is taken to mean more than a simple left-right model. 
A concrete example would be alternating emphasis, driven by information not available within the text \cite{ribeiro2015perceptual}.
Adding `external' inputs to distinguish between different possible realisations can be achieved with a multiple-regression HMM (MR-HMM) \cite{takashi2009hmm}.
Among the three changes that stuck out in \cite{watts2016hmms} 
as significant improvements, one was that of using expanded duration-dependent input features that allow for easy representation of optionality. 

\section{Conclusions}
\label{ch18-diagnose_conclusions}



We have seen that current approaches to both vocoding and statistical modelling limit the naturalness of contemporary parametric text-to-speech systems in a variety of ways. A natural question is then -- given what we have learned about these limitations from empirical data -- what are the most appealing research problems to pursue, in order to create improved parametric speech synthesisers?

For vocoding, sinusoidal signal representations appear capable of providing a low-level representation of the speech signal that maintains very high quality.
The question remains how to use this signal-fitting capacity for building a high-quality vocoder that allows faithful reconstruction of all perceived characteristics of the speech signal in the absence of any statistical modelling.

As more and more parts of speech synthesisers have been replaced by neural networks optimised via stochastic gradient descent (cf.\ \cite{takaki2015multiple}), the long-term goal of integrating aspects of the signal processing into the learned parts of the model has come to the forefront. The most extreme version of this agenda is statistical modelling in the raw waveform domain, which does not require a vocoder and may enable joint end-to-end optimisation of both signal processing and statistical modelling. This line of research recently saw a breakthrough in the form of the WaveNet paper from Google DeepMind \cite{oord2016wavenet}. A major downside, however, is the computational cost of the present WaveNet synthesiser, which simply is infeasible for practical applications. Reducing the computational load of these approaches whilst maintaining naturalness will no doubt be an important area of future research. In the meantime, non-parametric approaches to waveform generation continue to achieve the greatest segmental quality achievable in TTS applications, and it is not unreasonable to surmise that signal generators based on recorded speech will dominate applications for some time still.


In modelling, the advent of synthesisers based on deep and recurrent neural networks opened up a path to circumvent several long-standing limitations of decision-tree-based synthesis approaches. Most importantly, the use of neural networks reduced the amount of inappropriate conflation, and thus inappropriate averaging, performed by the learner: this includes both averaging across linguistic contexts, as well as along temporal positions, i.e., frame vs.\ state-level granularity.
In all likelihood, however, future research will identify new setups and approaches that better generalise from these contexts to the distribution of acoustic parameters or -- for waveform-level modelling -- the joint distribution of audio signal sample values.

There is another side to the conflation coin as well: instead of the all-or-nothing, binary averaging that decision tree methods perform, wherein datapoints in different leaves are treated entirely independently, deep learners might improve their models by learning from related but not directly relevant material, e.g, improve their modelling of one speech sound with the help of data from another, or better model one speaker by using additional speakers in the training data, as observed in \cite{fan2015multi} and also pursued in \cite{oord2016wavenet}. The argument is that the hierarchical design of deep learners allows useful information in otherwise unrelated linguistic contexts be extracted and processed into an abstracted form that is useful across context boundaries, a phenomenon dubbed the \emph{blessing of abstraction} \cite{tenenbaum2011grow}.
Historically, advances in speech synthesis have fed off the exponential growth of speech corpora sizes and computational resources seen in the last decades, and deep learning appears to benefit disproportionately much in the limit of very large amounts of data. The WaveNet paper, trained on 44 hours of data from more than 100 different speakers, is no exception to this rule. Pursuing methods that can adequately generalise from the vast amounts of unlabelled, multi-speaker, spontaneous speech material available all around us, even if it is not always directly relevant to the speech we want to synthesise, therefore appears to be another promising direction for future research.

In terms of generating output from distributions, we have learned that predicting the mean speech parameter sequence is not the route to naturalness, at least for speech parameterisations similar to STRAIGHT. From a theoretical perspective, only generation by sampling is -- essentially by definition -- certain to be able to achieve completely natural speech. Reaching high naturalness with sampled speech, however, places substantial demands on model accuracy: it is no longer sufficient to represent streams and individual parameters independently, but their dependences and collective behaviour must be accounted for in order to rise above the limitations identified in \cite{henter2014measuring}. In other words, more than one possible outcome must be represented well. 

The WaveNet paper \cite{oord2016wavenet} is probably the first published example of a synthesiser where random sampling has produced competitive-sounding speech, though it operates in the waveform domain.
Parametric speech synthesis is also likely to see additional efforts to perform accurate dependence modelling, but whether or not sampled parametric speech will come to surpass deterministically generated speech is hard to tell. This is especially true given that the limitations of mean-based generation do not necessarily preclude the existence of other deterministic methods capable of generating completely natural speech, if a suitable model is provided. For instance, it is presently unknown whether or not mean speech with GV-compensation (following, e.g., the methodology of \cite{nose2016efficient}) suffers the same upper limit as mean speech without such compensation. It is also not known what the upper limits on most likely parameter generation are -- or even whether or not that principle is subjectively preferable to mean speech output generation -- in models where the mean and mode do not coincide. These might be topics of future research.

The ultimate goal in speech technology research is not only to create synthesisers that sound natural, but enable technology that is natural to use. Considering the various flaws of contemporary speech synthesisers, the greatest practical issue may not be their segmental quality (which, WaveNet aside, can be made quite convincing in applications using hybrid synthesis from large speech corpora and by targetting the recorded prompts to the specific application domain, as evidenced by systems like Apple's Siri, Google Now, Microsoft Cortana, etc.), nor their intelligibility (which tends to be at ceiling in quiet conditions), but their awkward prosody and their ignorance of the communicative nature of speech and dialogue. It is often speculated, for instance by invoking references to the so called ``uncanny valley'' \cite{mori1970bukimi,moore2012bayesian}, that the adoption of TTS for practical tasks is limited not by segmental quality, but by the perceived unpleasantness of TTS systems. Furthermore, it is widely surmised that poor text and dialog/context understanding is a key bottleneck, and that more appropriate prosody and communication ability would be possible with improved linguistic features that better represent semantics and pragmatics.

An alternative, more practical approach may be to pursue better natural language processing, and from it derive more advanced and potentially more informative features for speech synthesis, just to see how much this can improve text-to-speech synthesis. Similar to the situation in speech processing, text and language processing have also seen substantial improvements from the adoption of deep learning, especially given the ease of acquiring truly gigantic text databases; NLP methods may be trained on much more text than any one human may read in their lifetime. If text and language processing continue to improve at the present rate, future developments will provide many interesting candidate techniques for integration into TTS. As a case in point of such technology transfer, one may consider NLP methods like \emph{word2vec} \cite{mikolov2013efficient} and related vector-space representations of text, or text and speech jointly \cite[for example]{rendel2016using,wang2016wordvector,watts2010letter,merritt2015deep}.


Whether through advanced features, through improved statistical sequence modelling, or from some other insight out of left field, accurate modelling and generation of prosody might very well be the final frontier in making synthetic speech pleasant, or at least palatable, to humans, in the context of an application. This might, in turn, be necessary for realising the transformative potential of TTS in places where speaking machines are scarcely more than a novelty, by finally providing humans with technological tools that are not merely powerful, but also natural to use.

\label{ch18end}






\renewcommand{\refname}{Bibliography}
\bibliography{nst}\label{refs}






\end{document}